\documentclass[article]{JHEP3}

\usepackage{amsmath}
\usepackage{amsfonts}
\usepackage{eufrak}
\usepackage{amssymb,bbold}
\usepackage{epsfig}
\usepackage{bm}

\def\be{\begin{equation}}
\def\ee{\end{equation}}

\def\a{\alpha}
\def\b{\beta}
\def\n{\nabla}
\def\t{\tau}

\def\v{\nu}
\def\m{\mu}
\def\s{\sigma}
\def\mbx{\mathbf{x}}
\def\o{\omega}

\def\pa{\partial}

\def\O{\Omega}
\def\e{\epsilon}
\def\G{\Gamma}
\def\md{\mathcal{D}}
\def\sq{\sqrt{2}}
\def\d{\delta}
\def\mr{\mathfrak{R}}
\def\to{\tilde{\o}_F}

\def\M{\mathcal{M}}
\def\td{\tilde{d}}
\newcommand{\eref}[1]{(\ref{#1})}
\newcommand{\alphah}{\ensuremath{\hat{\alpha}}}  
\newcommand{\betah}{\ensuremath{\hat{\beta}}}  
\newcommand{\ah}{\ensuremath{\hat{a}}}  
  
\newcommand{\ip}{\raise1pt\hbox{\large$\lrcorner$}\,}

\title{The general form of supersymmetric solutions of N=(1,0) U(1) and SU(2)
  gauged supergravities in six dimensions}
\author{Marco Cariglia\thanks{M.Cariglia@damtp.cam.ac.uk} and Ois\'{\i}n A. P. Mac Conamhna\thanks{O.A.P.MacConamhna@damtp.cam.ac.uk}\\DAMTP\\ Centre for Mathematical Sciences\\ University of Cambridge\\ Wilberforce Road, Cambridge CB3 0WA, UK.}

\preprint{DAMTP-2003-144} 

\abstract{We obtain necessary and sufficient conditions for a supersymmetric
field configuration in the N=(1,0) U(1) or SU(2) gauged
supergravities in six dimensions, and
impose the field equations on this general ansatz. It is found that
any supersymmetric solution is associated to an $SU(2)\ltimes
\mathbb{R}^4$ structure. The structure is characterized by a
null Killing vector 
which induces a natural 2+4 split of the six dimensional
spacetime. A suitable combination of the field equations implies that
the scalar curvature of the four dimensional Riemannian part, referred
to as the base, obeys a
second order differential equation. Bosonic fluxes introduce torsion terms
that deform the $SU(2)\ltimes\mathbb{R}^4$ structure away from a
covariantly constant one. The most general structure can be classified
in terms of its intrinsic torsion. For a large class of solutions the
gauge field strengths admit a simple geometrical interpretation: in
the U(1) theory the base is K\"{a}hler, and the gauge field strength is
the Ricci form; in the SU(2) theory, the gauge field strengths are
identified with the curvatures of the left hand spin bundle of the
base. We employ our general ansatz to construct new supersymmetric solutions; we show
that the U(1) theory admits a symmetric Cahen-Wallach$_4\times S^2$ solution
together with a compactifying pp-wave. The SU(2) theory admits a
black string, whose near horizon limit is $AdS_3\times S_3$. We also obtain
the Yang-Mills analogue of the Salam-Sezgin solution of the U(1)
theory, namely $R^{1,2}\times S^3$, where the $S^3$ is supported by a sphaleron. Finally we obtain the additional
constraints implied by 
enhanced supersymmetry, and discuss Penrose limits in the theories.} 
 
\keywords{gauged supergravities, G-structures}

\begin{document}
\section{Introduction}
Chiral N=(1,0) U(1) gauged supergravity \cite{sezgin1}, \cite{sezgin}
has received considerable attention both in the past \cite{salam},
\cite{hall} and more recently \cite{fer}-\cite{cve}, in part due
to phenomenological interest in the remarkable $\mathbb{R}^{1,3}\times
S^2$ solution found by Salam and Sezgin in \cite{salam}. Given this
recent interest, it is natural to attempt a classification of the
supersymmetric solutions of the theory, employing the powerful
techniques first used in \cite{tod} and developed in
\cite{gaunt}-\cite{klemm1}. We also, for the first time, apply this
technique to a non-abelian gauged supergravity, namely the N=(1,0)
gauged SU(2) theory in six dimensions. 

The strategy we use is to assume the
existence of at least one Killing spinor. Then we may construct from
that spinor a one form and a triplet of three forms, which satisfy
various algebraic and differential conditions which follow from the
Fierz identities and the Killing spinor equation. We exploit these,
and the supersymmetry variations of the other fermions in the theory,
to deduce the most general form of the bosonic fields compatible with
supersymmetry. The existence of a Killing spinor implies that most of
the equations of motion are satisfied identically. We impose the
remaining field equations on the general
supersymmetric ansatz. 

As we will see, the vector dual to the one form constructed from the
Killing spinor is both Killing and null. This induces a natural 2+4
split of the six dimensional spacetime. A combination of the field
equations and Bianchi identities of the three form field strength
present in the theory implies that the curvature of the  four dimensional
Riemannian part, or base, must obey a second order differential
equation. Solving this equation is the biggest obstacle we encounter;
disappointingly, in the U(1) theory we have been unable to find a base
which does not arise in known solutions which induces a non-singular
six dimensional metric. However, starting from ``known'' bases we have
been able to construct new six dimensional solutions. Furthermore this
general procedure for finding supersymmetric solutions yields
considerable geometrical insight into the form of the solutions. For
example, under a broad class of conditions (precisely specified below)
one may deduce that the base must be positive scalar curvature
K\"{a}hler, and that the gauge field strength is given by the Ricci
form of the base. 

Our results for the $SU(2)$ theory demonstrate that one may usefully
apply this general approach to nonabelian gauged supergravities. We
have been able to exploit the geometry of Killing spinors of the
theory, and in particular the fact that one can construct a triplet of
two forms which are anti self-dual on the base and (again for a broad class of
solutions) $SU(2)$ covariantly constant thereon. The existence of
these forms for this class of solutions allows the identification of
the gauge field strengths with the curvatures of the left-hand spin
bundle of the base. One might hope that something similar could be
achieved in other nonabelian gauged supergravities. 

The plan of the rest of this paper is as follows. In section 2 we give
a brief introduction to the theories we study, and in section 3 we
obtain the most general supersymmetric ansatz for each. We impose the
field equations on this ansatz in section 4, and in section 5 we
discuss intrinsic torsion. In section 6 we construct examples of
supersymmetric solutions solutions
of both theories. In section 7 
discuss solutions of the theories with enhanced supersymmetry, and in
section 8 we discuss the Penrose
limits of the theories. We conclude in section 9.    

\section{The supergravities}
We will work in mostly minus signature and adopt the conventions of
\cite{sezgin}.  All spinors of the theory are symplectic
Majorana, ie
\be
\chi^A=\e^{AB}\bar{\chi}_B^T,\;\;\;\bar{\chi}_A=(\chi^A)^{\dagger}\gamma_0.
\end{equation}
The $Sp(1)$ indices are raised and lowered as
\be
\chi^A=\e^{AB}\chi_B,\;\;\;\chi_A=\chi^B\e_{BA},\;\;\;\e^{12}=e_{12}=1.
\end{equation}
The field content of the $SU(2)$ theory is
follows: the gravity multiplet
$e^m_{\mu}$, $\psi_{\m L}^{A}$, $B^+_{\m\v}$, a tensor multiplet
$B^-_{\m\v}$, $\chi^A_R$, $\phi$ and an SU(2) gauge multiplet
$A^a_{\m}$, $\lambda^{aA}_L$. The subscripts denote the chiralities of
the fermions, and $^{\pm}$ means the potentials have self
and anti self dual field strengths. To translate the conventions of
\cite{salam} to those employed here one must change the signature of
the metric, switch from a Weyl spinor to a pair of symplectic Majorana
spinors, and also make the replacements
$(\kappa,\;\s,\;g)\rightarrow(1,\;\sq\phi,\;g/2)$.

Until section 6 we will treat the $U(1)$
and $SU(2)$ theories in tandem. Unless explicitly stated otherwise all
expressions given for the $SU(2)$ theory are valid for the $U(1)$ case
provided that two of the gauge potentials and the associated field
strengths are set to zero (and of course there is only one gaugino in
the $U(1)$ theory). We define
$G_3=dB_2+F^a_2\wedge
A^a_1-\frac{g}{6}f^{abc}A^a_1\wedge A^b_1\wedge A^c_1$ and
$F_2^a=dA_1^a+\frac{g}{2}f^{abc}A^b_1\wedge A^c_1$. The bosonic
Lagrangian of the theory is
\be
e^{-1}\mathcal{L}=-\frac{1}{4}R+\frac{1}{2}\pa_{\m}\phi\pa^{\m}\phi+\frac{1}{12}e^{2\sq\phi}G_{\m\v\s}G^{\m\v\s}-\frac{1}{4}e^{\sq\phi}F_{\m\v}^aF^{a\m\v}-\frac{ng^2}{8}e^{-\sq\phi},
\end{equation}
where $n=1$ for $U(1)$ and $n=3$ for $SU(2)$.
The fermion supersymmetry transformations are
\begin{eqnarray}
\label{psi}
\delta\psi^{A}_{\mu}=(\n_{\mu}-\frac{1}{4}e^{\sqrt{2}\phi}G^+_{\m\v\s}\Gamma^{\v\s})\e^A+gA^a_{\m}T^{aA}_{\;\;\;\;B}\e^B,\\\label{chi}\delta\chi^A=i(-\frac{1}{\sqrt{2}}\Gamma^{\mu}\partial_{\mu}\phi-\frac{e^{\sqrt{2}\phi}}{12}G^{-}_{\m\v\s}\Gamma^{\m\v\s})\e^{A},\\\label{lambda}\delta\lambda^{aA}=-\frac{e^{\phi/\sq}}{2\sqrt{2}}F^{a}_{\m\v}\Gamma^{\m\v}\e^A+\frac{e^{-\phi/\sqrt{2}}}{\sqrt{2}}gT^{aA}_{\;\;\;\;B}\epsilon^B,
\end{eqnarray}
where the $SU(2)$ generators $T^{aA}_{\;\;\;\;B}$ are given by
$T^{aA}_{\;\;\;\;B}=-\frac{i}{2}\s^{aA}_{\;\;\;\;B}$ and the
superscripts $^{\pm}$ denote the self and anti self dual
parts of $G$. The supersymmetry parameter $\e$ is left-handed, ie
$\G_7\e=-\e$, where
\be
\G_7\equiv\G_0\G_1...\G_5.
\end{equation}
The bosonic field equations and Bianchi identities are
\begin{eqnarray}
R_{\m\v}&=&2\pa_{\m}\phi\pa_{\v}\phi+e^{2\sq\phi}(G_{\m\s\lambda}G_{\v}^{\;\;\s\lambda}-\frac{1}{6}g_{\m\v}G_{\a\b\gamma}G^{\a\b\gamma})\nonumber\\&+&2e^{\sq\phi}(-F_{\m\a}^aF^{a\;\a}_{\v}+\frac{1}{8}g_{\m\v}F^a_{\a\b}F^{a\a\b})-\frac{ng^2}{8}e^{-\sq\phi}g_{\m\v},\\
\n^2\phi&=&\frac{1}{3\sq}e^{2\sq\phi}G_{\m\v\s}G^{\m\v\s}-\frac{1}{2\sq}e^{\sq\phi}F^a_{\m\v}F^{a\m\v}+\frac{ng^2}{4\sq}e^{-\sq\phi},\\dG&=&F^a\wedge
 F^a,\\d\star(e^{2\sq\phi}G)&=&0,\\dF^a&=&-g\e^{abc}A^b\wedge F^c,\\d\star(e^{\sq\phi}F^a)&=&-2e^{2\sq\phi}\star
G\wedge F^a-ge^{\sq\phi}\e^{abc}A^b\wedge\star F^c.\end{eqnarray}
Note that we have corrected the Einstein equation given in \cite{sezgin}.

\section{Necessary and sufficient conditions for supersymmetry}
Now we will implement the first part of our strategy and obtain the
general supersymmetric ansatz for our theories. Given a Killing spinor
$\e$ we may construct the nonzero bilinears
\begin{eqnarray}
V_{\m}\e^{AB}=\bar{\e}^A\G_{\m}\e^B,\\\O^{AB}_{\m\v\s}=\bar{\e}^A\G_{\m\v\s}\e^B.\end{eqnarray}
Forms of even degree vanish because $\e$ is chiral. The $\O^{AB}$ are
self dual and we define the real self dual forms $X^a$ 
($a=1,2,3$) by 
\begin{eqnarray}
\O^{11}&=&-\frac{1}{2}(X^2+iX^1),\\\O^{22}&=&-\frac{1}{2}(X^2-iX^1),\\\O^{12}&=&\O^{21}=\frac{1}{2}iX^3,\end{eqnarray}
so that
\be
\O^A_{\;\;\;B}=X^aT^{aA}_{\;\;\;\;B}.
\end{equation}
Note that this differs from \cite{reall}. Now the algebraic relations
satisfied by the bilinears imply the following condition given in
\cite{reall}
\be
V_{\m}V^{\m}=0.
\end{equation}
We introduce a null orthonormal basis 
\be
ds^2=2e^+e^--\delta_{ij}e^ie^j,
\end{equation}
where $e^+=V$, and we choose the orientation
\be
\e^{+-1234}=1.
\end{equation}
The algebraic relations also imply \cite{reall}
\be 
-\frac{1}{2}X^a=V\wedge I^a,
\end{equation}
where 
\be
I^a=\frac{1}{2}I^a_{ij}e^i\wedge e^j
\end{equation}
are anti-selfdual on the 4-d base with orientation
\be
\e^{ijkl}=\e^{+-ijkl},
\end{equation}
and obey
\be\label{Ialg}
(I^a)^i_{\;\;\;j}(I^b)^j_{\;\;\;k}=\e^{abc}(I^c)^i_{\;\;\;k}-\delta^{ab}\delta^i_k
\end{equation}
where the indices have been raised with $-\delta^{ij}$. It is also
shown in \cite{reall} that the Killing spinor must satisfy the projection
\be\label{proj}
\G^+\e=0.
\end{equation}

\subsection{Differential constraints}
Employing the Killing spinor equation, one may show that the covariant
derivative of $V$ is given by 
\be\label{nV}
\n_{\m}V_{\v}=e^{\sqrt{2}\phi}V^{\s}G^+_{\s\m\v},
\end{equation} 
so $V$ is Killing and $dV=2e^{\sq\phi}i_VG^+$, where $i_VK$ means $V$ contracted on
the first index of the form $K$. For $X^a$ we find
\be\label{O}
\n_{\a}X^a_{\m\v\s}=-g\e^{abc}A^b_{\a}X^c_{\m\v\s}+e^{\sqrt{2}{\phi}}(G^{+\t}_{\;\;\;\;\a\m}X^a_{\t\v\s}+G^{+\t}_{\;\;\;\;\a\v}X^a_{\t\s\m}+G^{+\t}_{\;\;\;\;\a\s}X^a_{\t\m\v}).
\end{equation}
We see that $\mathcal{L}_VX^a=0$ if we choose the gauge $i_VA^a=0$.

\subsection{$\delta\chi=0$}
Now we turn to the analysis of (\ref{chi}). On contracting $\d\chi=0$ with
$\bar{\e}^B$ we find that
\begin{eqnarray}
\label{dialaton}V^{\m}\partial_{\mu}\phi&=&0,\\X^{a\;\m\v\s}G^-_{\m\v\s}&=&0.
\end{eqnarray}
The duality relation for the gamma matrices is 
\be\label{gamma7}
\G^{\m_1...\m_n}=\frac{(-1)^{[n/2]}}{(6-n)!}\e^{\m_1...\m_n\v_1...\v_{6-n}}\G_{\v_1...\v_{6-n}}\G_7,
\end{equation}
with
\be
\e^{012345}=1,
\end{equation}
and in our null basis,
\begin{eqnarray}
\G_+&=&\frac{1}{\sq}(\G^0+\G^5),\\\G_-&=&\frac{1}{\sq}(\G^0-\G^5).
\end{eqnarray}
Contracting $\d\chi=0$ with $\bar{\e}^B\G_{\m\v}$ and employing the
duality relation we find
\begin{eqnarray}
V\wedge
d\phi&=&\sqrt{2}\;i_VG^-,\\G^-_{\t\rho[\m}X^{a\;\t\rho}_{\v]}&=&-X^{a\;\;\;\;\s}_{\m\v}\partial_{\s}(e^{-\sqrt{2}\phi}).
\end{eqnarray}
Together, these imply
\be\label{G-}
e^{\sq\phi}G^-=(1-\star)(-\frac{1}{\sq}V\wedge
e^-\wedge d\phi+\frac{1}{2}V\wedge K), \end{equation}
for some two form $K=\frac{1}{2}K_{ij}e^i\wedge e^j$. The anti
self-duality  of $G^-$ implies that $K$ is self dual with respect to
$-\delta_{ij}$.

\subsection{$\delta\lambda^{aA}=0$}
In analysing the consequences of this equation it is convenient to
note the following for any product of gamma matrices $A$:
\begin{eqnarray}
\frac{1}{2}\bar{\e}^AF_{\m\v}^a[\G^{\m\v},A]\e^B+ge^{-\sqrt{2}\phi}(T^{aA}_{\;\;\;\;C}\delta^B_D+T^{aB}_{\;\;\;\;D}\delta^A_C)\bar{\e}^CA\e^D&=&0,\\\frac{1}{2}\bar{\e}^AF_{\m\v}^a\{\G^{\m\v},A\}\e^B+ge^{-\sqrt{2}\phi}(T^{aA}_{\;\;\;\;C}\delta^B_D-T^{aB}_{\;\;\;\;D}\delta^A_C)\bar{\e}^CA\e^D&=&0.
\end{eqnarray}
Now, with $A=\G_{\s}$, we find
\begin{eqnarray}
i_VF&=&0,\\
F^{a\;\m\v}X^b_{\s\m\v}&=&2ge^{-\sq\phi}\delta^{ab}V_{\s}.
\end{eqnarray}
With $A=\G_{\m\v\s}$ and using (\ref{gamma7}) we find
\begin{eqnarray}
6X^{b\;\;\;\;\v}_{[\s\rho}F^a_{\t]\v}&=&-ge^{-\sqrt{2}\phi}\e^{abc}X^c_{\s\rho\t},\\V\wedge F^a+\star(V\wedge F^a)&=&\frac{1}{4}ge^{-\sqrt{2}\phi}X^a.
\end{eqnarray}
These imply that 
\be\label{F}
F^a=V\wedge \omega^a_F+\tilde{F}^a-\frac{g}{4}e^{-\sqrt{2}\phi}I^a,
\end{equation}
where $\omega^a_F=\omega^a_{Fi}e^i$, and
$\tilde{F}^a=\frac{1}{2}\tilde{F}^a_{ij}e^i\wedge e^j$ is self dual
with respect to $-\delta_{ij}$. Now one may show that
\be
\mathcal{L}_VG=\mathcal{L}_VF^a=0,
\end{equation}
and in the gauge chosen above, also that $\mathcal{L}_VA^a=0$.

\subsection{Sufficient conditions for supersymmetry}
Now we show that the necessary conditions for supersymmetry we have
derived are (when supplemented by further projections from the gauge
multiplet) also sufficient. We begin by analysing $\d\chi$. In the
basis chosen above, given the projection (\ref{proj}), and noting that
$\G^+$ and $\G^-$ do not anticommute, it
is immediate that if $G^-$ is given by (\ref{G-}), then
$\delta\chi$ reduces to
\be
\d\chi^A=i\Big(-\frac{1}{2\sq}\G^i\partial_i\phi-\frac{1}{12\sq}\e_{ijkl}\G^{ijk}\partial^l\phi\Big)\e^A.
\end{equation}
Now we note that on the base,
\be
\G^{i_1...i_n}=\frac{(-1)^{[n/2]}}{(4-n)!}\e^{i_1...i_nj_i...j_{4-n}}\G_{j_1...j_{4-n}}\G_{\star},
\end{equation}
where $\G_{\star}\equiv\G_{1234}$. Since $\G_7\e=-\e$ and
$\G^+\e=0$ we
have that $\G_{\star}\e=\e$. Thus $\d\chi=0$.
 
Next we turn to $\delta\lambda^a$. If $F^a$ is given by (\ref{F}) then again because of the projection (\ref{proj}), the
$V\wedge \omega_F$ term gives zero. The
self duality of $\tilde{F}^a$ with respect to $-\delta_{ij}$ implies that
$\tilde{F}^a_{ij}\G^{ij}\e=0$.
Thus we find that
\be
\delta\lambda^{aA}=\frac{ge^{-\phi/\sqrt{2}}}{\sqrt{2}}\Big(\frac{1}{8}I^a_{ij}\G^{ij}\delta^A_B+T^{aA}_{\;\;\;\;B}\Big)\e^B.
\label{eq:gaugino_variation}
\end{equation}
At first it might appear that the vanishing of $\delta\lambda^a$ requires three further projections
on $\e$ thus breaking all supersymmetry. However we note that
\be
\Big[\Big(\frac{1}{8}I^a_{ij}\G^{ij}+T^{a}\Big), 
\Big(\frac{1}{8}I^b_{kl}\G^{kl}+T^b\Big)\Big]^A_{\;\;B}=\e^{abc}\Big(\frac{1}{8}I^c_{ij}\G^{ij}\delta^A_B+T^{cA}_{\;\;\;\;B}\Big),
\end{equation}
so any two projections imply the third and generically one
supersymmetry is left unbroken. In the $U(1)$ theory we have only one
gaugino so only one further projection, and thus generically two
supersymmetries are unbroken.

Finally we turn to the Killing spinor equation. All terms involving a
 $\G^-$ vanish because
(\ref{nV}) is equivalent to 
\be\label{o=G}
\omega_{\m\v-}=e^{\sqrt{2}\phi}G^{(+)}_{\m\v-}.
\end{equation}
Then all further terms involving a $\G^+$ vanish because of
(\ref{proj}). Therefore we find that
\be
\delta\psi_{\m}=(\partial_{\m}+\frac{1}{4}\omega_{\m
  ij}\G^{ij}+gA_{\m}^aT^a-\frac{e^{\sqrt{2}\phi}}{4}G_{\m
  ij}^{(+)}\G^{ij})\e.
\end{equation}
Now the terms involving the parts of $\omega_{\m ij}$ and $G_{\m ij}$
which are self dual in the indices $i,j$ vanish as above. To analyse
the anti-self dual part we note that (\ref{O}) implies that
\be\label{dI}
\n_{\a}I^a_{ij}+g\e^{abc}A^b_{\a}I^c_{ij}=e^{\sqrt{2}\phi}(G^{(+)k}_{\a
  i}I^a_{kj}-G^{(+)k}_{\a
  j}I^a_{ki}),
\end{equation}
and, if we choose the basis so that the components of the $I^a$ are
constants, this becomes 
\be\label{ptak}
(\omega_{\a ij}-e^{\sqrt{2}\phi}G^{(+)}_{\a
  ij})^{\tilde{-}}=\frac{g}{2}A_{\a}^aI^a_{ij},
\end{equation}
where $\tilde{-}$ denotes the anti-self dual projection in
$i,j$. Hence the variation of the gravitino reduces to 
\begin{eqnarray}
\delta\psi_{\m}&=&\Big(\partial_{\m}+gA^a_{\m}\Big(\frac{1}{8}I^a_{ij}\G^{ij}+T^{a}\Big)\Big)\e\nonumber\\&=&\partial_{\mu}\e,
\end{eqnarray}
if $\e$ obeys the projections required for the vanishing of
$\delta\lambda$. Thus in this basis, given the algebraic and
differential constraints on $V$ and $X$, and the form of the fields
given above, the Killing spinor equation is satisfied by any constant
spinor satisfying the requisite projections. Thus we have derived
necessary and sufficient conditions for supersymmetry. 

\section{The field equations}
In this section we will impose the field equations on our general
ansatz. We
introduce the local coordinates of \cite{reall}:
\be
ds^2=2H^{-1}(du+\beta_mdx^m)\Big(dv+\o_ndx^n+\frac{\mathcal{F}}{2}(du+\b_ndx^n)\Big)-Hh_{mn}dx^mdx^n,
\label{eq:metric_ansatz} 
\end{equation}
with
\begin{eqnarray}
e^+&=&H^{-1}(du+\beta_mdx^m),\nonumber\\e^-&=&dv+\o_mdx^m+\frac{\mathcal{FH}}{2}e^+.
\end{eqnarray}
As vectors,
\begin{eqnarray}
e^+&=&\frac{\pa}{\pa v},\\e^-&=&H\Big(\frac{\pa}{\pa
  u}-\frac{\mathcal{F}}{2}\frac{\pa}{\pa v}\Big),
\end{eqnarray}
and $H,\;\mathcal{F},\;\o,\;\b$ and $h_{mn}$ depend on $u$ and $x$ but
  not on $v$, which is the affine parameter along the null geodesics
  to which the Killing vector $e^+$ is tangent.

In what follows we will employ some further notation of
\cite{reall}. Specifically, let $\Phi$ be a form defined on the base
with
\be
\Phi=\frac{1}{p!}\Phi_{i_1...i_p}(u,x)dx^{i_1}\wedge...\wedge dx^{i_p},
\end{equation}
and let
\be
\tilde{d}\Phi=\frac{1}{(p+1)!}(p+1)\frac{\pa}{\pa
  x^{[q}}\Phi_{i_1...i_p]}dx^q\wedge dx^{i_1}\wedge...\wedge dx^{i_p},
\end{equation}
and define the operator $\md$ as
\be
\md\Phi=\tilde{d}\Phi-\b\wedge\dot{\Phi},
\end{equation}
where $\dot{\Phi}$ denotes the Lie derivative with respect to
$\frac{\pa}{\pa u}$. Then we have
\be
d\Phi=\md\Phi+He^+\wedge\dot{\Phi}.
\end{equation}
Further note that
\begin{eqnarray}
de^+&=&H^{-1}\md\b+e^+\wedge(H^{-1}\md
H+\dot{\b}),\\de^-&=&\md\o+\frac{\mathcal{F}}{2}\md\b+He^+\wedge(\dot{\o}+\frac{\mathcal{F}}{2}\dot{\b}-\frac{1}{2}\md\mathcal{F}).
\end{eqnarray}
We define $J^a=H^{-1}I^a$. They obey 
\be
(J^a)^i_{\;\;j}(J^b)^j_{\;\;k}=-\e^{abc}(J^c)^{i}_{\;\;k}-\d^{ab}\d^j_{\;\;k},
\end{equation}
where, here and henceforth, indices on the base are raised with
$h^{mn}$. If we define
 \be
A^a=A^a_+(u,x)e^++\tilde{A}^a,\;\;\tilde{A}^a=A^a_i(u,x)e^i,
\end{equation}
then equation (\ref{O}) implies that
\be\label{J}
\tilde{d}J^a+g\e^{abc}\tilde{A}^b\wedge J^c=\partial_u(\b\wedge J^a).
\end{equation} 
Antisymmetrising the $k$th component of (\ref{dI}), employing the
expression for $G^+$ given below, and comparing with
(\ref{J}), we see that $\b\wedge\pa_u(HJ^a)=0$.

We still have the freedom to make a $v$-independent gauge
transformation on the 1-form potential, while preserving the condition
$i_VA^a=0$. Since $A_+^a$ is independent of $v$, we may exploit this
freedom to set $A_+^a=0$ which we do in what follows. Also we note
that (\ref{dialaton}) implies that $\phi=\phi(u,x)$.

\subsection{The three form}
Equations (\ref{G-}), (\ref{o=G}) and
the $+$ component of (\ref{dI}) imply that 
\begin{eqnarray}
e^{\sqrt{2}\phi}G&=&\frac{1}{2}\star_4(\md
H+H\dot{\b}-\sq H\md\phi)+e^+\wedge(-H\psi-\frac{1}{2}(\md\o)^-+K)\nonumber\\&+&e^-\wedge\frac{\md\b}{2H}-\frac{1}{2}e^+\wedge
e^-\wedge(H^{-1}\md H+\dot{\b}+\sq\md\phi),
\end{eqnarray}
where
\be
\psi=\frac{H}{16}\e^{abc}J^{aij}\dot{J}^b_{ij}J^c,
\end{equation}
and also
\be
\md\b=\star_4\md\b,
\end{equation}
with $\star_4$ the Hodge dual on the base with metric $h_{mn}$. If we
write $\d\psi_{\m}=D_{\m}\e,\;\d\chi=\Delta_G\epsilon$, then we have
the integrability condition
\begin{eqnarray}\label{int1}
i\G^{\m}[D_{\m},\Delta_G]\e&=&\frac{1}{\sq}\Big(\n^2\phi-\frac{1}{3\sq}e^{2\sq\phi}G_{\m\v\s}G^{\m\v\s}+\frac{1}{2\sq}e^{\sq\phi}F^a_{\m\v}F^{a\m\v}\nonumber\\&-&\frac{ng^2}{4\sq}e^{-\sq\phi}\Big)\e+\frac{1}{4}e^{-\sq\phi}(\star
d\star(e^{2\sq\phi}G))_{\m\v}\G^{\m\v}\e\nonumber\\&+&\frac{1}{48}e^{\sq\phi}(dG-F^a\wedge
F^a)_{\m\v\s\tau}\G^{\m\v\s\tau}\e-\frac{i}{3}G_{\m\v\s}\G^{\m\v\s}\d\chi\nonumber\\&-&\Big(\frac{1}{2\sq}e^{\phi/\sq}F^a_{\m\v}\G^{\m\v}+\frac{g}{\sq}e^{-\phi/\sq}T^a\Big)\d\lambda^a=0.
\end{eqnarray}
Thus the existence of a Killing spinor means that if we impose the
three form Bianchi identities and field equations, the dilaton
field  equation is automatically satisfied. Defining 
\be
\mathcal{G}=\frac{1}{2H}((\md \o)^++\frac{1}{2}\mathcal{F}\md \b),
\end{equation}
the $+ij$ components of the three form field
equations and the Bianchi identities give
\begin{eqnarray}
\md(H^{-1}e^{\sq\phi}(K-H\mathcal{G}-H\psi))&+&\frac{1}{2}\pa_u\star_4(\md(He^{\sq\phi})+e^{\sq\phi}H\dot{\b})\nonumber\\\label{tree1}-H^{-1}e^{\sq\phi}\dot{\b}\wedge(K-H\mathcal{G}-H\psi)&=&0
\end{eqnarray}
and
\begin{eqnarray}
-\md(H^{-1}e^{-\sq\phi}(K+H\mathcal{G}-H\psi))&+&\frac{1}{2}\pa_u\star_4(\md(He^{-\sq\phi})+e^{-\sq\phi}H\dot{\b})\nonumber\\\label{tree2}+H^{-1}e^{-\sq\phi}\dot{\b}\wedge(K+H\mathcal{G}-H\psi)&=&2H^{-1}\o^a_F\wedge(\tilde{F}^a-\frac{g}{4}He^{-\sq\phi}J^a),
\end{eqnarray}
while the $ijk$ components are
\begin{eqnarray}\label{dil}
\frac{1}{2}e^{-\sq\phi}\md\star_4\Big(\md(e^{\sq\phi}H)&+&e^{\sq\phi}H\dot{\b}\Big)=H^{-1}(K-H\mathcal{G})\wedge\md\b,\\\label{goat}\frac{1}{2}e^{\sq\phi}\md\star_4\Big(\md(e^{-\sq\phi}H)&+&e^{-\sq\phi}H\dot{\b}\Big)+H^{-1}(K+H\mathcal{G})\wedge\md\b\nonumber\\&=&e^{\sq\phi}\tilde{F}^a\wedge\tilde{F}^a+\frac{H^2}{16}g^2e^{-\sq\phi}J^a\wedge J^a.
\end{eqnarray}

\subsection{The two forms}
Writing $\d\lambda^a=\Delta_F^a\e$, we obtain the following
integrability condition
\begin{eqnarray}\label{int2}
\sq\G^{\m}[D_{\m},\Delta_F^a]\e&=&-\frac{1}{6}e^{\phi/\sq}(dF^a+g\e^{abc}A^b\wedge
  F^c)_{\m\v\s}\G^{\m\v\s}\e\nonumber\\&-&e^{-\phi/\sq}P_{\m}^a\G^{\m}\e-\G^{\m}\pa_{\m}\phi\d\lambda^a-ie^{\phi/\sq}F^a_{\m\v}\G^{\m\v}\d\chi\nonumber\\&+&\sq
  i[\Delta_G,\Delta_F^a]\e-\sq g\G^{\m}\e^{abc}A^b_{\m}\d\lambda^c,
\end{eqnarray}
with
\be
P_{\m}^a=\Big(\star\Big[d\star(e^{\sq\phi}F^a)+2e^{2\sq\phi}\star G\wedge
F^a+ge^{\sq\phi}\e^{abc}A^b\wedge \star F^c\Big]\Big)_{\m}.
\end{equation}
The two form field equations are $P^a_{\m}=0$. If we impose the
Bianchi identities we have 
$\G^{\m}P_{\m}^a\e=0$. Acting with $\bar{\e}$ and $\G^{\v}P^a_{\v}$ we
find $P_-^a=P^a_{\m}P^{a\m}=0$. Hence the existence of a Killing
spinor together with the Bianchi identities implies that all except
the $+$ component of the field equations are automatically satisfied.
Imposing $F^a=dA^a+\frac{g}{2}\e^{abc}A^b\wedge A^c$, we find that
\begin{eqnarray}
\tilde{F}^a-\frac{g}{4}e^{-\sq\phi}HJ^a&=&\md\tilde{A}^a+\frac{g}{2}\e^{abc}\tilde{A}^b\wedge\tilde{A}^c,\\\label{defom}\o^a_F&=&H\dot{\tilde{A}}^a.
\end{eqnarray}
We may invert the $\a=k$ components of equation (\ref{dI}) to solve
for $A$ and obtain
\be
A^a_i=\frac{1}{8g}\Big[-\e^{abc}J^{bjk}\tilde{\n}_iJ^c_{jk}-\b_i\e^{abc}J^{bjk}\dot{J}^c_{ij}+4\dot{\b}_jJ^{aj}_{\;\;\;i}+2\b_jJ^{a}_{ki}\dot{h}^{jk}\Big],
\end{equation}
where $\tilde{\n}$ denotes the Levi-Civita connection on the base with
metric $h_{mn}$. The $+$ component of the field equation gives an
equation for $\o_F^a$ and reads

\begin{eqnarray}
\md(H\star_4\o^a_F)&=&-2\sq\md\phi\wedge(H\star_4\o^a_F)+2\tilde{F}^a\wedge(K-H\mathcal{G})\nonumber\\\label{omega}&-&\frac{g}{2}He^{-\sq\phi}J^a\wedge(H\psi+(\md\o)^-)-gH\e^{abc}\tilde{A}^b\wedge\star_4\o^c_F.
\end{eqnarray}

\subsection{The Einstein equations}
It may be shown with considerable effort that the integrability
condition for the Killing spinor equation is
\begin{eqnarray}\label{kill}
\G^{\v}[D_{\m},D_{\v}]\e&=&\frac{1}{2}E_{\m\v}\G^{\v}\e+\frac{e^{\sq\phi}}{96}(dG-F^a\wedge
    F^a)_{\a\b\gamma\delta}\G^{\a\b\gamma\delta}\G_{\m}\e\nonumber\\&+&\frac{e^{-\sq\phi}}{8}(\star
    d\star(e^{2\sq\phi}G))_{\a\b}\G^{\a\b}\G_{\m}\e\nonumber\\&+&\sq
    e^{\phi/\sq}F^a_{\m\v}\G^{\v}\d\lambda^a-i(\sq\pa_{\m}\phi-\frac{1}{12}G^-_{\a\b\gamma}\G^{\a\b\gamma}\G_{\m})\d\chi\nonumber\\&-&\frac{1}{2\sq}\G_{\m}\Big(\frac{e^{\phi/\sq}}{2}F^a_{\a\b}\G^{\a\b}+ge^{-\phi/\sq}T^a\Big)\d\lambda^a,
\end{eqnarray}
where
\begin{eqnarray}
E_{\m\v}&=&-R_{\m\v}+2\pa_{\m}\phi\pa_{\v}\phi+e^{2\sq\phi}(G_{\m\s\lambda}G_{\v}^{\;\;\s\lambda}-\frac{1}{6}g_{\m\v}G_{\a\b\gamma}G^{\a\b\gamma})\nonumber\\&+&2e^{\sq\phi}(-F_{\m\a}^aF^{a\;\a}_{\v}+\frac{1}{8}g_{\m\v}F^a_{\a\b}F^{a\a\b})-\frac{ng^2}{8}e^{-\sq\phi}g_{\m\v},
\end{eqnarray}
and the Einstein equations are
\be
E_{\m\v}=0.
\end{equation}
Given the vanishing of the supersymmetry variations of $\chi$ and
$\lambda^a$, and that $G$ satisfies its field equation and Bianchi
identity, we see upon acting on (\ref{kill}) with $\bar{\e}$ and
$E_{\m\s}\G^{\s}$ that all except the $++$ component of the Einstein
equations are 
implied by the integrability of the Killing spinor equation. The $++$
component is
\begin{eqnarray}
\star_4\md(\star_4[\dot{\o}&+&\frac{1}{2}\mathcal{F}\dot{\b}-\frac{1}{2}\md\mathcal{F}])=\frac{1}{2}Hh^{mn}\partial_u^2(Hh_{mn})+\frac{1}{4}\partial_u(Hh^{mn})\partial_u(Hh_{mn})\nonumber\\&-&2\dot{\b}_m(\dot{\o}+\frac{1}{2}\mathcal{F}\dot{\b}-\frac{1}{2}\md\mathcal{F})^m\nonumber-\frac{1}{2}H^{-2}(\md\o+\frac{1}{2}\mathcal{F}\md\b)^2\nonumber\\&+&2H^{-2}(K+H\psi+\frac{1}{2}(\md\o)^-)^2+2H^2\dot{\phi}^2+2H^{-1}e^{\sq\phi}\o_F^{ai}\o_{Fi}^a
\end{eqnarray}
where for a two form $M$,
\be
\M^2=\frac{1}{2}M_{ij}M^{ij}.
\end{equation}

\section{Fluxes and intrinsic torsion} 
The geometrical structure we have studied so far is associated to a
chiral spinor and is given by $SU(2)\ltimes \mathbb{R}^4$. $SU(2)$
corresponds to rotations in the base, while $\mathbb{R}^4$ to null
rotations that leave the Killing spinor invariant. The objects that
define such a structure are the Killing vector $K$ and the triplet
of anti-selfdual forms $I^a$. 
The presence of fluxes in the Killing equation implies that such
objects are not covariantly constant with respect to the Levi-Civita
connection, but are covariantly constant with respect to a connection
with torsion. One can then classify the various inequivalent
spacetimes by studying their intrinsic torsion. The intrinsic torsion
can be roughly described as the obstruction to finding a torsion free
connection on the spacetime. Given a pair of connections, their
different is always a tensor field  $\alpha^\lambda_{\mu\nu}$, with one
covariant and two 
contravariant indices. The difference of their associated torsion
tensors is then given by
$\alpha^\lambda_{\nu\mu}-\alpha^\lambda_{\mu\nu}$. It is clear then
that it is possible to obtain a new connection with zero torsion if and
only if the original torsion field could be written as
$\alpha^\lambda_{\mu\nu}-\alpha^\lambda_{\nu\mu}$ for some tensor
field $\alpha$. Given a spacetime with a torsion tensor
$T^\lambda_{\mu\nu} =T^\lambda_{[\mu\nu ]}$, its intrinsic torsion is
defined as the projection of $T$ on the quotient space obtained via
this subtraction procedure.  
 
The gaugino transformation law implies a set of projections on
$\epsilon$ given by \eref{eq:gaugino_variation}. Plugging these into
\eref{psi} one can rewrite this as 
\be 
	\delta\psi_\mu^A = \nabla_\mu^T \, \epsilon^A = 
\left[ \partial_\mu   + \frac{1}{4} \left( \omega_{\mu\alphah\betah} +
T_{\mu\alphah\betah} \right) \Gamma^{\alphah\betah} \right] \epsilon^A   , 
\ee 
where $\alphah$, $\betah$ are flat $6D$ indices, and
$T_{\lambda\mu_1\mu_2}$ is a tensor given by 
\be 
	T_{\lambda\mu_1\mu_2} = - \left(
e^{\sqrt{2}\Phi}G^+_{\lambda\mu_1\mu_2} + \frac{g}{2} A_\lambda^a \,
I^a_{ij} \, 
\delta^i_{\mu_1} \, \delta^j_{\mu_2} \right) . \label{eq:tensor} 
\ee 
$\lambda$ is the index associated to differentiation, while $\mu_1$,
$\mu_2$ are cotangent space indices associated to the group action of
$SO(6)$ on tensors. Eq.\eref{eq:tensor} is clearly antisymmetric in
the indices $\mu_1$, $\mu_2$, but in general it is not with respect to
$\lambda$ and $\mu_2$. This latter symmetry is that required in order
to interpret $T$ as a torsion tensor. Now, notice that equations
\eref{nV}, \eref{dI} can be rewritten 
as 
\begin{eqnarray} 
	\nabla_\mu^T V_\nu &=& 0 , \nonumber \\ 
	\nabla_\mu^T I^a_{ij} &=& 0 . 
\end{eqnarray}
This tells us that there exists a suitable connection such that the
$SU(2)\ltimes \mathbb{R}^4$ structure can be seen as covariantly
constant, as in the case without fluxes. The overall effect of fluxes
is that of deforming the geometry, while keeping the same geometric
structure. When \eref{eq:tensor} is antisymmetric in
the indices $\lambda$, $\mu_2$ then we can think of it as a torsion
tensor associated to the Levi-Civita connection.  
If there is a part
symmetric in $\lambda$, $\mu_2$ then instead one has to consider a
connection which is more general than the Levi-Civita one.

In principle one can study in detail the intrinsic torsion in six
dimensions. However, as we are going to see later, for most of the
applications one is mostly interested in fully understanding the geometry on
the four-dimensional base. Therefore we are going to restrict
ourselves to systematically describe the geometry on the base. All the
information that is not included in such geometry is encoded in
$\dot{\tilde{A}}^a$ and in the components of $G^+$ laying along $u$ and $v$
directions. These basically correspond to derivatives of $H$ and of
the twisting parameters $\beta$, $\omega$ appearing in the mertic
\eref{eq:metric_ansatz}.  
 
On the base there is an $Sp(1)$ structure with torsion, see below for
a definition. In order  to
calculate it project equation \eref{dI} on the base and get 
\be 
	\tilde{\nabla}_i\, J^a_{jk} + g \epsilon^{abc} A_i^b
J_{jk}^c - \beta_i \dot{J}^a_{jk} - \dot{\beta}_{[j}\,J^a_{k]i} +
h_{i[j}\,\dot{\beta}^m\,J^a_{k]m} = 0 .  \label{eq:djay_base2}
\ee 
Where indices are raised with the metric $h^{ij}$. 
This corresponds to the vanishing of a covariant derivative with a
tensorial part given by 
\be 
 \tilde{T}_{ijk} = \frac{g}{2} A_i^a \, J^a_{jk} + \beta_i \tilde{e}^{\ah}_j
\dot{\tilde{e}}_{\ah k} -\left( h_{i[j}\dot{\beta}_{k]}\right)   . 
\label{eq:tensor2}
\ee 
Again there is explicit antisymmetry with respect to the indices $j$,
$k$, but not with respect to $i$ and $k$, and therefore the same
remarks made for eq.\eref{eq:tensor} apply. In four dimensions a
rather general class of manifolds such that the tensor
\eref{eq:tensor2} can be directly seen as a torsion tensor is given by
Hyper K\"{a}hler manifolds with torsion (HKT), which are those such
that \eref{eq:tensor2} is completely antisymmetric.

An $Sp(1)$ structure can be shown to be equivalent to an $SU(2)$
one. In our case the $SU(2)$ structure can be obtained once we choose
one of the three $J^a$ to be a complex structure. Choose $J^1$ for
example and define 
\begin{eqnarray} 
	J &=& J^1 , \nonumber \\ 
	\Omega &=& J^2 + i J^3 . 
\end{eqnarray} 
It is not difficult to show that they define an $SU(2)$ structure,
that satisfies the defining equations  
\begin{eqnarray} 
	J \wedge \Omega &=& 0 , \nonumber \\ 
	\Omega \wedge \overline{\Omega} &=& 2 J \wedge J . 
\end{eqnarray} 
$SU(2)$ structures are completely understood. Their  intrinsic torsion
can be decomposed into invariant representations, called modules. In
our case there are three such modules, given by  
\begin{eqnarray} 
	\mathcal{W}_2 &=& \frac{1}{4}  \left( \Omega \ip d\Omega +
	\overline{\Omega} \ip d\overline{\Omega} \right) , \nonumber
	\\ 
	\mathcal{W}_4 &=& J \ip d J ,\nonumber
        \\
        \mathcal{W}_5&=&\frac{1}{4}\left(\O\ip d
	\overline{\O}+\overline{\O}\ip d\O\right),\nonumber
\end{eqnarray}
where we define the contraction $\ip$ as $(\omega_2 \ip \omega_3)_k := 1/2
\omega_2^{ij} \, \omega_{3 \, ijk}$. Now, calculate $dJ$, $d\Omega$
using (4.14) and $\b\wedge\pa_u(HJ)=0$. Equivalently one can antisymmetrize
(5.4) and obtain the same, as a check of
consistency. The result is 
\begin{eqnarray} 
	\mathcal{W}_2 &=& \frac{g}{2} \left( A^2\,J^2 - A^3\,J^3 \right)
, \nonumber 
	\\ 
	\mathcal{W}_4 &=& -g \left( A^2\,J^2 + A^3\,J^3
	\right)+H\pa_u(H^{-1}\b) ,\nonumber
\\
\mathcal{W}_5&=&-\frac{g}{2}(A^2J^2+A^3J^3)-gA^1J^1+H\pa_u(H^{-1}\b).  \nonumber
\end{eqnarray}
When $\mathcal{W}_2 =0$ the structure is integrable, and this corresponds to $A^2\ip J^2 = A^3 \ip J^3$. When both
$\mathcal{W}_2$ and $\mathcal{W}_4$ are zero instead the manifold is
K\"ahler. This corresponds, for $\pa_u(H^{-1}\beta)=0$, to
$A^2=0=A^3$ that is, the $U(1)$ theory. This case will be studied in
detail in the next section.

\section{Supersymmetric solutions}
When either $\b=0$ or the full system is independent of $u$, the general
problem simplifies considerably. We 
then have
\begin{eqnarray}\label{JJ}
\tilde{\n}_iJ^a_{jk}&=&-g\e^{abc}A^b_iJ^c_{jk},\nonumber\\A^a_i&=&-\frac{1}{8g}\e^{abc}J^{bij}\tilde{\n}_iJ^c_{jk}.
\end{eqnarray}
Note that for the $U(1)$ theory, this implies that the base is
K\"{a}hler, since then one of the $J$s is covariantly constant on the base. Now
using (\ref{JJ}) 
and 
\be
\tilde{F}^a-\frac{g}{4}e^{-\sq\phi}HJ^a=\tilde{d}\tilde{A}^a+\frac{g}{2}\e^{abc}\tilde{A}^b\wedge\tilde{A}^c,
\end{equation}
we find that 
\be\label{lick}
(\tilde{F}^a-\frac{g}{4}e^{-\sq\phi}HJ^a)_{ij}=-\frac{1}{2g}J^{amn}R_{mnij}\equiv-\frac{1}{g}\mr^a_{ij},
\end{equation}
with $R_{ijkl}$ the Riemann tensor on the base with metric
$h_{mn}$. Contracting with $J^a$ we find
\be\label{riff}
e^{-\sq\phi}H=\frac{1}{ng^2}\mr^a_{ij}J^{aij}\equiv\frac{1}{g^2}\mr.
\end{equation}
As we shall see below, $\mr$ is proportional to the scalar curvature
of the base, with positive constant of proportionality. Thus, since we
are taking $H$ positive, (this amounts to the our choice of the
signature of spacetime) the base must have positive scalar
curvature. If we in fact assume that $\b=0$ (that is, we seek what we
shall refer to as non-twisting solutions), then on employing
(\ref{lick}) and (\ref{riff}), we find that 
(\ref{dil}) and 
(\ref{goat}) become
\begin{eqnarray}\label{harm}
\tilde{\n}^2(e^{2\sq\phi}\mr)&=&0,\\\label{base}\tilde{\n}^2\mr&=&\frac{n}{2}\mr^2-\mr^a_{ij}\mr^{aij}.
\end{eqnarray}
Thus we see that we must choose a (possibly $u$-dependent) base satisfying
(\ref{base}). Then we choose a harmonic function, $f$, on the base,
and we have 
\begin{eqnarray}
e^{2\sq\phi}&=&\frac{f}{\mr},\\H&=&\frac{1}{g^2}\sqrt{f\mr}.
\end{eqnarray}
We now employ the gauge freedom present in our choice of coordinates
to set $\o=0$. Then $\mathcal{G}=0$ also, and on rescaling
$\o^a_F$ according to
\begin{eqnarray}
\o^a_F&=&e^{-\sq\phi}\tilde{\o}^a_F,
\end{eqnarray}
and defining the two form $Y$ as
\be
Y=K-H\psi,
\end{equation}
 equations (\ref{tree1}),
(\ref{tree2}) and (\ref{omega}) reduce to the coupled system
\begin{eqnarray}
\tilde{d}\Big(\frac{1}{\mr}\star_4Y\Big)&=&-\frac{1}{2g^4}\pa_u(\star_4\tilde{d}f),\\\tilde{d}\Big(\frac{1}{f}Y\Big)&=&\frac{1}{2g^4}\pa_u(\star_4\tilde{d}\mr)+\frac{2}{gf}\tilde{\o}^a_F\wedge\mr^a,\\\tilde{d}(\star_4f\tilde{\o}^a_F)&=&-\frac{2gf}{\mr}\mr^a\wedge
\star_4Y+g\e^{abc}\tilde{A}^b\wedge
f\star_4\tilde{\o}^c_F.
\end{eqnarray}
The $++$ component of the Einstein equation is
\begin{eqnarray}
\frac{1}{2}\tilde{\n}^2\mathcal{F}&=&\frac{1}{g^4}\bigg[\mr^2\bigg(\pa_u\sqrt{\frac{f}{\mr}}\bigg)^2+\frac{1}{2}\sqrt{f\mr}h^{mn}\pa_u^2(\sqrt{f\mr}h_{mn})\nonumber\\&+&\frac{1}{4}\pa_u(\sqrt{f\mr}h^{mn})\pa_u(\sqrt{f\mr}h_{mn})\bigg]+\frac{2g^4}{f\mr}Y^2+\frac{2g^2}{f}(\to^{a})^2.
\end{eqnarray}
We will now consider in turn the $U(1)$ and $SU(2)$ theories.

\subsection{Non-twisting solutions of the $U(1)$ theory}
We take the $U(1)$ generator to be $T^1$. Then the base is K\"{a}hler
  with gauge-invariant K\"{a}hler form $J^1$ and Ricci form
  $\mr^1$. The Ricci form obeys
\be
\mr^1_{ij}=R_{ik}J^{1k}_{\;\;\;j},
\end{equation}
where $R_{ij}$ is the Ricci tensor of the base. Thus we have
$R=\mr=J^{1ij}\mr^1_{ij}$, with $R$ the Ricci scalar of  the
base. Equation (\ref{base}) becomes
\be\label{u1base}
\tilde{\n}^2R=\frac{1}{2}R^2-R^{ij}R_{ij}.
\end{equation}

\subsubsection{Dyonic string}
Let us take the base to be of the form
\be
ds^2=a^{-2}dr^2+\frac{r^2}{4}((\s_R^1)^2+(\s_R^2)^2)+a^2\frac{r^2}{4}(\s_R^3)^2,
\end{equation}
where the left-invariant 1-forms $\s_R^a$ obey
$d\s_R^a=1/2\e^{abc}\s_R^b\wedge\s_R^c$. Taking an orthonormal basis to be
given by 
\be
e^1=\frac{r}{2}\s_R^1,\:\;e^2=\frac{r}{2}\s_R^2,\;\;e^3=a\frac{r}{2}\s_R^3,\;\;e^4=a^{-1}dr,
\end{equation}
the K\"{a}hler form is $J^1=e^1\wedge e^2-e^3\wedge e^4$, and the
vierbein components of the Ricci tensor are
\begin{eqnarray}
R_{11}=R_{22}&=&\frac{4}{r^2}(1-a^2),\\R_{33}=R_{44}&=&0.
\end{eqnarray}
Thus the base has everywhere positive scalar curvature if $a<1$.
Now we may recover the dyonic string solutions of
\cite{pope}. Let us take the full solution to be independent of $u$,
and take $K=\mathcal{F}=0$. Then
choosing 
\be
f=\frac{g^4}{8(1-a^2)}\Big(Q_1+\frac{Q_2}{r^2}\Big),
\end{equation}
the full solution is completely determined; it is
\begin{eqnarray}
ds^2&=&\frac{r^2}{\sqrt{r^2Q_1+Q_2}}\eta_{\m\v}dx^{\m}dx^{\v}\nonumber\\&-&\frac{1}{r}\sqrt{Q_1+\frac{Q_2}{r^2}}\Big(a^{-2}dr^2+\frac{r^2}{4}[(\s_R^1)^2+(\s_R^2)^2+a^2(\s_R^3)^2]\Big),\\F&=&-\frac{1}{g}(1-a^2)\s_R^1\wedge\s_R^2,\;\;\;e^{2\sq\phi}=\frac{g^4(r^2Q_1+Q_2)}{64(1-a^2)^2},
\end{eqnarray}
and $G$ is determined by $H$ and $\phi$.

\subsubsection{Base a product of two manifolds}
Next we take the base to be a product of two real oriented Riemannian
manifolds, $\mathcal{B}=\mathcal{M}_1\times\mathcal{M}_2$. We take
$J^1$ to be $Vol_1-Vol_2$, where $Vol_i$ the volume form of
$\mathcal{M}_i$. Then with $R_i$ the scalar curvature of
$\mathcal{M}_i$, (\ref{u1base}) gives
\be
\tilde{\n}^2(R_1+R_2)=R_1R_2.
\end{equation}
The Salam-Sezgin model \cite{salam} is an example of such a solution
with $\M_1=\mathbb{R}^2$, $\M_2=S^2$. More generally,
$\mathbb{R}^2\times\mathcal{M}_2$ where $\mathcal{M}_2$ has everywhere
positive harmonic scalar curvature is also an allowed solution. However we will
now constuct a u-dependent generalisation of the Salam-Sezgin
model, by allowing the radius of the $S^2$ to depend on $u$. We will
find that the four dimensional part of the metric takes the form of a
pp-wave. We take the base to be of the form
\be
ds_4^2=dx^2+dy^2+t^2(u)(d\theta^2+\sin^2\theta d\phi^2).
\end{equation}
We take $H=1$, and so we have
\begin{eqnarray}
R&=&\frac{2}{t^2},\\e^{\sq\phi}&=&\frac{g^2t^2}{2}.
\end{eqnarray}
Defining
\be
e^1=dx,\;\;e^2=dy,\;\;e^3=td\theta,\;\;e^4=t\sin\theta d\phi,
\end{equation}
we choose the $J^i$ to be
\be\label{ji}
J^1=e^1\wedge e^2 -e^3\wedge e^4,\;\;J^2=e^1\wedge e^4-e^2\wedge
e^3,\;\;J^3=e^1\wedge e^3+e^2\wedge e^4,
\end{equation}
so $\psi=0$, $\dot{\tilde{A}}=0$ and 
\be
\tilde{F}-\frac{g}{4}e^{-\sq\phi}J^1=\frac{1}{gt^2}e^3\wedge e^4.
\end{equation}
We choose $K=0$, so that $G=0$, and the Einstein equation
reduces to
\be
\tilde{\n}^2\mathcal{F}=\frac{8\dot{t}^2}{t^2}+\frac{4\ddot{t}}{t}.
\end{equation}
This is solved by 
\be
\mathcal{F}=\Big(2\frac{\dot{t}^2}{t^2}+\frac{\ddot{t}}{t}\Big)\mbx^2.
\end{equation}
If we choose $t=\cosh u$, the full metric is given by
\be
ds^2=2du(dv+\frac{1}{2}(1+2\tanh^2u)\mbx^2du)-d\mbx^2_2-\cosh^2ud\Omega^2_2.
\end{equation}
With this choice of $t$, we have a sort of dynamical supersymmetric
compactification; the spacetime is effectively four dimensional for
small $|u|$, and decompactifies for large $|u|$.

\subsubsection{Cahen-Wallach$_4\times S^2$}
We conclude our discussion of non-twisting solutions of the $U(1)$
theory by presenting the 
symmetric space $CW_4\times S^2$ as a solution which preserves one
quarter supersymmetry, as well as maximal four dimensional symmetry
(this was not obtained by the authors of \cite{boss} as they only
looked for
solutions with four dimensional Poincar\'{e}, de Sitter or Anti de
Sitter symmetry). Specifically, we take the base to be
$\mathbb{R}^2\times S^2$, with u-independent metric
\be
ds^2=d\mbx^2_2+a^2(d\theta^2+\sin^2\theta d\phi^2).
\end{equation}
With the obvious choice of orthonormal basis, the K\"{a}hler form is
$J^1=e^1\wedge e^2-e^3\wedge e^4$. We choose $H=1$, so $f=\frac{a^2g^4}{2}$.
Now, however, we will take $J^2$ and $J^3$ to be u-dependent. Defining
\be
L^2=e^1\wedge e^4-e^2\wedge e^3,\;\;\;L^3=e^1\wedge e^3+e^2\wedge e^4,
\end{equation}
we make the following choice for $J^2$ and $J^3$:
\begin{eqnarray}
J^2&=&\cos2uL^2+\sin2uL^3,\\J^3&=&-\sin2uL^2+\cos2uL^3.
\end{eqnarray}
Hence, 
\be
\psi=-J^1.
\end{equation}
Our system of equations reduces
to
\begin{eqnarray}
\label{O1}\tilde{d}(K-J^1)&=&0,\\\label{O2}\tilde{d}(K+J^1)&=&0,\\\label{O3}(K-J^1)\wedge F&=&0,\\\label{O4}\tilde{\n}\mathcal{F}&=&4(K^2+(J^1)^2).
\end{eqnarray}
Equations (\ref{O1})-(\ref{O3}) are solved by 
\be
K=e^1\wedge e^2+e^3\wedge e^4,
\end{equation}
and (\ref{O4}) is
\be
\tilde{\n}^2\mathcal{F}=16,
\end{equation}
which is solved by
\be
\mathcal{F}=4\mbx^2.
\end{equation}
The full solution is then 
\begin{eqnarray}
ds^2&=&2du(dv+2\mbx^2du)-d\mbx_2^2-a^2d\O_2^2,\\G&=&\frac{4}{a^2g^2}du\wedge
dx^1\wedge dx^2,\\F&=&\frac{1}{ga^2}e^3\wedge
e^4,\\e^{\sq\phi}&=&\frac{a^2g^2}{2}. 
\end{eqnarray}
Since $K\neq0$ this solution preserves one quarter supersymmetry, and
to our knowledge it has not been given previously.

\subsection{Non-twisting solutions of the $SU(2)$ theory}
In analysing
the $SU(2)$ theory it is convenient to find expressions 
for $R_{ij}$ and $R$ in terms of $\mr^a$ and $\mr$. Defining 
\be
\hat{\n}_iJ_{jk}^a=\tilde{\n}_iJ_{jk}^a+g\e^{abc}A^b_iJ^c_{jk},
\end{equation}
we have
\be
[\hat{\n}_i,\hat{\n}_j]J_{mn}=0,
\end{equation}
which implies that
\be
R_{ijmn}=\frac{1}{3}(J^{ak}_iJ^{a\:l}_jR_{klmn}+2J^a_{ij}\mr^a_{mn}),
\end{equation}
on employing (\ref{lick}). From this we find
\be
J^{ak}_iR_{kj}=\mr^a_{ij}-\e^{abc}J^{bk}_i\mr^c_{kj},
\end{equation}
and hence that
\begin{eqnarray}
\mr&=&\frac{1}{3}R,\\\mr^a_{ij}\mr^{aij}&=&R_{ij}R^{ij}+\e^{abc}\mr^{aj}_i\mr^b_{jk}J^{cik}.
\end{eqnarray}
On employing (\ref{lick}) again we find
\be
\mr^a_{ij}\mr^{aij}=R_{ij}R^{ij}-\frac{1}{6}R^2,
\end{equation}
and thus equation (\ref{base}) is
\be\label{su2base}
\tilde{\n}^2R=R^2-3R_{ij}R^{ij},
\end{equation}
for the $SU(2)$ theory. This equation is precisely the
requirement that the Weyl anomaly for $\mathcal{N}=4$ $U(N)$ SYM
vanishes on the base. Whether this is a mere coincidence or has some
deeper significance is unclear. 

In fact, the Yang-Mills field strengths are precisely the curvatures
of the left-hand spin bundle of the base (see eg \cite{pope2}). The
base must again have positive scalar curvature. Furthermore for
supersymmetry we must only allow bases such that the anti-self dual
parts of the curvatures of
the left hand spin bundle are of the specific form 
\be\label{qquat}
\mr^{a(-)}=\frac{R}{12}J^a,
\end{equation}
from (\ref{lick}). In
particular, K\"{a}hler bases are excluded, since then the anti
selfdual parts of the curvatures are all proportional to the
K\"{a}hler form, and this is inconsistent with supersymmetry. We note
in passing that (\ref{qquat}) is reminiscent of a quaternionic
K\"{a}hler manifold in higher dimensions - the quaternionic K\"{a}hler
condition is vacuous in four dimensions - though in the case at hand,
the selfdual parts of the curvatures may (and indeed for a solution of
(\ref{su2base}) must) be nonzero.  

\subsubsection{Base $\mathbb{R}\times\mathcal{M}^3$} 
Equation (\ref{su2base}) admits solutions of the form
$\mathbb{R}\times\mathcal{M}^3$, with $\mathcal{M}^3$ Einstein. Choosing
$\mathcal{M}^3$ to be an $S^3$ with round metric (more generally
we could have the lens space $S^3/\mathbb{Z}_p$), and
writing the base metric as
\be
ds^2=dx^2+\frac{a^2}{4}\Big((\s_R^1)^2+(\s_R^2)^2+(\s_R^3)^2\Big),
\end{equation}
we satisfy the constraint (\ref{qquat}). We choose
the $J^i$ as in (\ref{ji}) with
\be
e^1=dx,\;\;e^2=\frac{a}{2}\s_R^1,\;\;e^3=\frac{a}{2}\s_R^3,\;\;e^4=\frac{a}{2}\s_R^2,  
\end{equation} 
and we also choose $H=1$, $f=\frac{a^2g^4}{2}$, so that
\be
e^{\sq\phi}=\frac{a^2g^2}{2}.
\end{equation}
Then $G=\o_F=0$, and taking $\mathcal{F}=0$, the solution is 
\begin{eqnarray}
ds^2&=&dt^2-d\mbx^2_2-a^2d\Omega^2_3\nonumber,\\F^a&=&-\frac{1}{8g}\e^{abc}\s^b_R\wedge\s^c_R,
\end{eqnarray}
which is the Yang-Mills analogue of the Salam-Sezgin model; the $S^3$
is supported by a sphaleron (the Yang-Mills potentials are given by
$A^a=-(2g)^{-1}\s^a_R$) and the solution
preserves 1/4 supersymmetry, which is the maximum possible in the
$SU(2)$ theory. We may
easily find the Yang-Mills analogue of the compactifying pp-wave found
in the U(1) theory; it is
\begin{eqnarray}
ds^2&=&2du(dv+(3+4\tanh^2u)x^2du)-dx^2-\cosh^2ud\Omega^2_3\nonumber,\\F^a&=&-\frac{1}{8g}\e^{abc}\s_b\wedge\s_c,\;\;\;e^{\sq\phi}=\frac{g^2}{2}\cosh^2u,\;\;\;G=0.
\end{eqnarray}

\subsubsection{Dyomeronic black string and $AdS_3\times S^3$}
Now we will show that the $SU(2)$ theory admits a black string
solution with dyonic three form charges and a meron on the transverse
space. It is straightforward to verify that the metric
\be
ds^2=dr^2+\frac{a^2r^2}{4}\s_R^a\s_R^a
\end{equation}
is a (singular) positive scalar curvature solution of (\ref{su2base})
and (\ref{qquat}) when
$a^2<1$. Again we  choose
the $J^i$ as in (\ref{ji}) with
\be
e^1=dr,\;\;e^2=\frac{ar}{2}\s_R^1,\;\;e^3=\frac{ar}{2}\s_R^3,\;\;e^4=\frac{ar}{2}\s_R^2.
\end{equation}
Then, the nonzero vielbein components of the Riemann tensor of the base
are
\be
R_{3434}=R_{2323}=R_{2424}=\frac{1}{a^2r^2}(1-a^2),
\end{equation}
and the scalar curvature is
\be
R=\frac{6}{a^2r^2}(1-a^2).
\end{equation}
Let us take $f$ to be given by
\be
f=\frac{a^2g^4}{2(1-a^2)}\Big(Q_1+\frac{Q_2}{r^2}\Big),
\end{equation}
where (for a nonsingular six dimensional solution) we require $Q_1$, $Q_2>0$. Then 
\begin{eqnarray}
H&=&\frac{1}{r}\sqrt{Q_1+\frac{Q_2}{r^2}},\\e^{2\sq\phi}&=&\frac{a^4g^4}{4(1-a^2)^2}(r^2Q_1+Q_2).
\end{eqnarray}
Let us take $K=\mathcal{F}=0$. Then the $F^a$ are given by
\be
F^a=-\frac{(1-a^2)}{8g}\e^{abc}\s^b_R\wedge\s^c_R.
\end{equation}
The metric is
\be
ds^2=\frac{r}{\sqrt{Q_1+\frac{Q_2}{r^2}}}\eta_{\m\v}dx^{\m}dx^{\v}-\frac{1}{r}\sqrt{Q_1+\frac{Q_2}{r^2}}(dr^2+\frac{a^2r^2}{4}\s_R^a\s_R^a),
\end{equation}
and $G$ is determined by $H$. The metric is everywhere nonsingular,
and has a horizon at $r=0$. In the 
near horizon limit $Q_1\rightarrow0$, defining $r=\sqrt{Q_2}u^{-1}$, the metric becomes
\be
ds^2=\sqrt{Q_2}\Big(\frac{1}{u^2}(\eta_{\m\v}dx^{\m}dx^{\v}-du^2)-\frac{a^2}{4}\s_R^a\s_R^a\Big),
\end{equation}
which is $AdS_3\times S^3$, the radius of curvature of the $AdS$
factor being greater than that of the $S^3$. We may recover the
$R^{1,2}\times S^3$ solution by setting the 3-form flux to zero in the
limit $Q_2\rightarrow\infty$, $a^2\rightarrow0$, $\sqrt{Q_2}a^2$ fixed.

\subsection{$u$-independent solutions}
In the interests of completeness we will write down the system
of equations determining the general $u$-independent
solution. Equations (\ref{JJ})-(\ref{riff}) remain valid, and the
remaining equations reduce to
\begin{eqnarray}
\label{big1}\td\star_4\td(\mr e^{2\sq\phi})&=&\frac{2g^4}{\mr}(K-H\mathcal{G})\wedge\td\b,\\
\label{big2}\td\star_4\td
\mr&=&2\mr^a\wedge\mr^a\nonumber\\&-&\frac{2g^4}{e^{2\sq\phi}\mr}(K+H\mathcal{G})\wedge\td\b,\\
\label{big3}\td\Big(\frac{1}{\mr}[K-H\mathcal{G}]\Big)&=&0,\\
\label{big4}\td\Big(\frac{e^{-2\sq\phi}}{\mr}[K+H\mathcal{G}]\Big)&=&0,\\
\label{big5}\mr^a\wedge(K-H\mathcal{G}+(\td\o)^-)&=&0,\\
\label{big6}-\frac{1}{2}\star_4\td\star_4\td\mathcal{F}&=&2H^{-2}(K^2-H^2\mathcal{G}^2).
\end{eqnarray}
We note that these may be solved as follows: take
$\o=K=\mathcal{F}=0$, the base to be given by a
solution of (\ref{u1base}) or (\ref{su2base}) as appropriate, and $\b$
to be a form with self-dual field strength on the base but otherwise
arbitrary. We have been unable to find any other nonsingular solutions.

\section{Solutions with enhanced supersymmetry}
In this section we will obtain further constraints on the bosonic
fields implied by enhanced supersymmetry - that is, we consider solutions
preserving one half or one quarter supersymmetry in the $U(1)$ and
$SU(2)$ theories respectively. Given two linearly independent Killing
spinors $\e$, $\e^{\prime}$, with associated Killing vectors $V$,
$V^{\prime}$, we always have
\be
V_{\m}\G^{\m}\e=V_{\m}^{\prime}\G^{\m}\e^{\prime}=0.
\end{equation}
However there is no reason why we should have
$V_{\m}\G^{\m}\e^{\prime}=0$ or $V_{\m}^{\prime}\G^{\m}\e=0$ (see 5.1
of \cite{gaunt} for further discussion of this point). Thus we will
relax the condition $\G^+\e=0$ in this section, though we may of
course still employ our general ansatz.  

\subsection{U(1) solutions preserving one half supersymmetry}
Let us define
\be
F_B=\tilde{F}-\frac{g}{4}He^{-\sq\phi}J^1.
\end{equation}
Now we know from our general ansatz that
\be
F=e^+\wedge\o_F+F_B.
\end{equation}
Then, the vanishing of $\d\lambda$ implies that
\be
(\o_{Fi}\G^{+i}+F_{Bij}\G^{ij}+2ge^{-2\sq\phi}T^1)\e=0.
\end{equation}
Employing $(\G^+)^2=0$, $\G^+\e\neq0$ then gives
\begin{eqnarray}
\label{FB}F_{Bij}\G^{ij}\e&=&-2ge^{-2\sq\phi}T^1\e,\\\o_{Fi}\G^i\e&=&0.
\end{eqnarray}
We impose the projection (\ref{FB}), and require that this is the only
algebraic constraint on the Killing spinor. Hence we must impose
$\o_F=0$. Next we note that
\be
(F_{Bij}\G^{ij}+2ge^{-\sq\phi}T^1)\d\lambda=0,
\end{equation}
so 
\be 
(-F_{Bij}F_{Bkl}\G^{ijkl}+2F_{Bij}F_B^{ij}-g^2e^{-2\sq\phi})\e=0,
\end{equation}
and hence
\begin{eqnarray}
\label{fwedge}F_B\wedge F_B&=&0,\\\label{fcont}F_{Bij}F_B^{ij}&=&\frac{g^2}{2}e^{-\sq\phi},
\end{eqnarray}
and so $F_B$ is a decomposable two form.

For the vanishing of $\delta\chi$ to imply no further restrictions on
the Killing spinor, we clearly must have $\phi=$ constant,
$G^-=0$. Then given (\ref{fwedge}), the three form field equations are
implied by the Bianchi identities, and (\ref{fcont}) is equivalent to
the dilaton field equation. Requiring $[D_{\m},\Delta_F]\e=0$ implies 
\be
(\n_{\m}F_{\v\s}+2e^{\sq\phi}G^{+\;\;\;\t}_{\m\v}F_{\s\t})\G^{\v\s}\e=0,
\end{equation}
and hence that
\begin{equation}
\n_{\m}F_{\v\s}=-2e^{\sq\phi}G^{+\;\;\;\t}_{\m[\v}F_{\s]\t},
\end{equation}
which (for constant dilaton) implies the field equations for
$F$. Given the two form Bianchi identities we also find
\be
G^{+\;\;\;\t}_{[\m\v}F_{\s]\t}=0.
\end{equation}
Next $[D_{\m},D_{\v}]\e=0$ together with $\delta\lambda=0$ implies
\begin{eqnarray}\label{int4}
(R_{\m\v\s\t}&-&e^{\sq\phi}\n_{\m}G^+_{\v\s\t}+e^{\sq\phi}\n_{\v}G^{+}_{\m\s\t}\nonumber\\&-&2e^{2\sq\phi}G^{+\;\;\;\rho}_{\m\s}G^+_{\v\t\rho}+2e^{\sq\phi}F_{\m\v}F_{\s\t})\G^{\s\t}\e=0,
\end{eqnarray}
and hence the expression in parentheses must vanish. Antisymmetrising this
expression on $\v$, $\s$ and $\t$ and employing (\ref{fwedge})
together with the
Bianchi identity and self-duality of $G^+$, we find
\be
\n G^+=0.
\end{equation}
Thus $G^+$ is parallel with respect to the Levi-Civita connection. We
therefore find for the Riemann tensor that
\be\label{Rie}
R_{\m\v\s\t}=-2(e^{2\sq\phi}G^{+\;\;\;\a}_{\m[\s}G^+_{\t]\v\a}+e^{\sq\phi}F_{\m\v}F_{\s\t}).
\end{equation}
Given equation (\ref{fcont}) and that $\phi=$ constant, $G^-=0$, this
implies that all the Einstein equations are satisfied.

We will not attempt a complete analysis of solutions with one half
supersymmetry, imposing these additional constraints on our general
ansatz. Rather we shall show the uniqueness of the Salam-Sezgin 
vacuum among one half supersymmetric solutions with vanishing three
form flux, and give a brief discussion of a special class of solutions. 

\subsubsection{$G^+=0$}
When $G^+=0$, the 6 dimensional Riemann tensor is parallel with respect to the
Levi-Cicita connection and the geometry is
locally symmetric. Since $F$ is decomposable and its only nonvanishing
components are spacelike, (\ref{Rie}) implies that the solution is
locally isometric to $\mathbb{R}^{1,3}\times \mathcal{M}_2$, where
$\mathcal{M}_2$ is a symmetric Riemannian two manifold. Equation
(\ref{fcont}) then implies that $\mathcal{M}_2=S^2$. Thus any
solution with vanishing three form flux which preserves one half
supersymmetry is locally isometric to $\mathbb{R}^{1,3}\times S^2$. If we
assume in addition simple connectedness then any such solution is in
fact isometric to $\mathbb{R}^{1,3}\times S^2$. 

\subsubsection{One half supersymmetric solutions with K\"{a}hler base} 
When the base is K\"{a}hler, we know that $F$ equals the
Ricci form of the base. The decomposability of the Ricci form implies that the
Ricci tensor of the base has two zero eigenvalues. Then equation
(\ref{u1base}) implies that the other pair of eigenvalues (which are
necessarily equal since the base is K\"{a}hler) is harmonic; in order
that the base have positive scalar curvature they must also be
everywhere positive. If the nonzero pair of eigenvalues are in fact
constant, and we assume that the base is compact, then according to
\cite{apost} the base is locally the product $T^2\times S^2$. When the
nonzero pair of eigenvalues is not constant, we can offer no such
general result. Instead we will consider a specific example; consider
a base equipped with the metric
\be\label{Kbase}
ds^2=\frac{dr^2}{W^2(r)}+\frac{r^2}{4}((\s_R^1)^2+(\s_R^2)^2+W^2(r)(\s_R^3)^2).
\end{equation}
Choosing the vierbeins as
\be
e^1=\frac{r}{2}\s_R^1,\;\;\;e^2=\frac{r}{2}\s_R^2,\;\;\;e^3=\frac{rW}{2}\s_R^3,\;\;\;e^4=W^{-1}dr,
\end{equation}
we find the following for the vierbein components of the Ricci tensor:
\begin{eqnarray}
R_{11}=R_{22}&=&-\frac{2WW^{\prime}}{r}-\frac{4W^2}{r^2}+\frac{4}{r^2},\\R_{33}=R_{44}&=&-WW^{\prime\prime}-(W^{\prime})^2-\frac{5WW^{\prime}}{r}.
\end{eqnarray}
Now requiring $R_{11}=0$ gives $W^2=1-ar^{-4}$, namely
Eguchi-Hanson, which is Ricci flat and thus not an allowed
base. Therefore, we impose $R_{33}=0$. On employing
$R_{33}=0$, we find that $R^{\prime}=-2r^{-1}R$, hence
$R=ar^{-2}$. Next,
imposing $\n^2R=0$, or
\be
R^{\prime}=\frac{b}{r^3W^2},
\end{equation}
we get $W=$ constant, which solves $R_{33}=0$. Finally for a positive
scalar curvature Riemannian metric, we must have $0<W<1$. This is
therefore the unique base of the form (\ref{Kbase}) which induces a
one half supersymmetric solution. This base was employed in the
construction of the $U(1)$ dyonic string, which is one quarter
supersymmetric, as it does not have constant
dilaton. To get a one half supersymmetric solution, we take $f=cR$ to
obtain the $AdS_3$ times a squashed $S^3$ solution of \cite{pope},
which arises as the near horizon limit of the dyonic string.

\subsection{$SU(2)$ solutions preserving one quarter supersymmetry}
To obtain solutions of the $SU(2)$ theory preserving one quarter
supersymetry, we must again relax the condition
that $\G^+\e=0$. As before, this implies that $\o_F^a=0$. We must then
impose any two of the projections $\d\lambda^a=\d\lambda^b=0$, together
with $[\Delta_{F^a},\Delta_{F^b}]\e\sim\e^{abc}\Delta_{F^c}\e$. Thus
we find 
\be\label{alg}
F^{a\;\;\;k}_{B[i}F^b_{Bj]k}=\frac{g}{4}e^{-\sq\phi}\e^{abc}F^{c}_{Bij}.
\end{equation}
Now we also have
\be
(F^a_{Bij}\G^{ij}+2ge^{-\sq\phi}T^a)\d\lambda^b=0,
\end{equation}
or
\be
\Big(\frac{1}{2}F_B^{aij}F_{Bkl}^b([\G_{ij},\G^{kl}]-\{\G_{ij},\G^{kl}\})+g^2e^{-2\sq\phi}(2\e^{abc}T^c-\d^{ab})\Big)\e=0. 
\end{equation}
Using (\ref{alg}), this becomes
\be
(-F^a_{Bij}F^b_{Bkl}\G^{ijkl}+2F_{Bij}^aF_{B}^{bij}-g^2e^{-2\sq\phi}\d^{ab})\e=0,\end{equation}  
and thus
\begin{eqnarray}
\label{suf}F^a_B\wedge
F^b_B&=&0,\\\label{suc}F^a_{Bij}F^{bij}_B&=&\frac{g^2}{2}e^{-2\sq\phi}\d^{ab}.
\end{eqnarray}
By arguments identical to those employed in the analysis of the $U(1)$
theory, we find that the following constraints, together with
(\ref{alg}), (\ref{suf}) and (\ref{suc}), are implied by the
requirement of one quarter supersymmetry:
\begin{eqnarray}
\phi&=&const,\\G^-&=&0,\\\label{aaa}R_{\m\v\s\t}&=&-2(e^{2\sq\phi}G^{+\;\;\;\a}_{\m[\s}G^+_{\t]\v\a}+e^{\sq\phi}F^a_{\m\v}F^a_{\s\t}),\\\n
G^+&=&0,\\\label{bbb}
\n_{\m}F^a_{\v\s}+2g\e^{abc}A^b_{\m}F^c_{\v\s}&=&-2e^{\sq\phi}G^{+\;\;\;\t}_{\m[\v}F^a_{\s]\t},
\end{eqnarray}
and given the Bianchi identities of the forms, all field equations are
satisfied. We could impose these additional
constraints on our general ansatz to more fully characterise solutions
preserving one quarter supersymmetry. However we will merely observe that
they are satisfied by our $AdS_3\times S^3$ and
$\mathbb{R}^{1,2}\times S^3$ solutions. The latter is the unique
solution with vanishing three form flux which preserves one quarter
supersymmetry. To see this, we note that when $G=0$, (\ref{aaa}) and
(\ref{bbb}) imply that $R_{\m\v\s\t}$ is parallel with respect to
$\n$, and hence the geometry is locally symmetric. Further since the $F^a$
only have nonzero components on the base, the solution is necessarily
locally isometric to $\mathbb{R}^{1,1}\times \mathcal{M}_4$, where
$\mathcal{M}_4$ is a symmetric positive scalar curvature non
K\"{a}hler Reimannian manifold. By direct inspection of equation
(\ref{su2base}), we see that the only possibility for the base is
$\mathbb{R}\times S^3$. Thus a solution with vanishing three form
flux which preserves one quarter supersymmetry is necessarily locally
isometric to $\mathbb{R}^{1,2}\times S^3$, with isometry if we assume
simple connectedness.

\section{Penrose Limits}
In this section we will define the Penrose limits of the
$U(1)$ and $SU(2)$ theories and employ our definition to derive
nonabelian pp-wave solutions, closely following \cite{G},
\cite{Fg}. In the neighbourhood of a segment of a null geodesic
containing no conjugate points, we may introduce local null coordinates
$U$, $V$, $X^i$ such that the metric takes the form
\be
ds^2=dV\Big(dU+\a dV+\rho_idX^i\Big)-C_{ij}dX^idX^j.
\end{equation}
We may also choose a gauge such that the one and two form potentials
satisfy
\be
A^a_U=B_{U\a}=0.
\end{equation}
As usual, we introduce a positive real constant $\O$ and rescale the
coordinates: 
\be
U=u,\;\;\;V=\O^2v,\;\;\;X^i=\O x^i.
\end{equation}
We thus obtain an $\O$-dependent family of fields $g_{\m\v}(\O)$,
$\phi(\O)$, $A^a_{\m}(\O)$, $B_{\m\v}(\O)$. We define the familiar
rescaled fields, distinguished by an overbar:
\begin{eqnarray}
\bar{g}_{\m\v}(\O)&=&\O^{-2}g_{\m\v}(\O),\\\bar{\phi}(\O)&=&\phi(\O),\\\bar{A}^a_{\m}(\O)&=&\O^{-1}A^a_{\m}(\O),\\\bar{B}_{\m\v}(\O)&=&\O^{-2}B_{\m\v}(\O).
\end{eqnarray}
However under these rescalings neither the Lagrangian nor the
supersymmetry variations of the theory 
transform homogeneously; the terms which do not transform
appropriately are the nonlinear $gA\wedge A$ terms in the Yang-Mills
field strengths, the $gA\wedge A\wedge A$ terms in the three form
field strength and the $ng^2e^{-\sq\phi}$ term in the Lagrangian
(which is also present in 
the $U(1)$ theory). Thus if we want a well defined Penrose limit which
takes supersymmetric solutions into supersymmetric solutions we must
rescale the gauge coupling constant according to
\be
\bar{g}=\O g,
\end{equation}
so that the Yang-Mills and three form field strengths are rescaled according to
\begin{eqnarray}
\bar{F}^a(\O)&=&\O^{-1}F^a(\O),\\
\bar{G}(\O)&=&\O^{-2}G(\O),
\end{eqnarray}
and then the Lagrangian and supersymmetry transformations transform
homogeneously. The Penrose limit consists of taking the limit of the
barred fields (and coupling) as $\O\rightarrow0$; in particular, we
have $\bar{g}\rightarrow0$, so we will obtain pp-wave solutions of the
ungauged theory. The solutions of the SU(2) theory are non-abelian
pp-waves of the same kind as those described in $4$-dimensions in
\cite{non-abelian}. 

\subsection{An example: the Penrose limit of $AdS_3\times S^3$} 
Defining $R_1$, $R_2$ to be the radii of curvature of the $AdS_3$ and
$S^3$ factors respectively, and $a=R_2/R_1$, let us write our
$AdS_3\times S^3$ solution in the following form: 
\begin{eqnarray}
R_1^{-2}ds^2&=&dt^2-\sin^2t\Big[\frac{dr^2}{1+r^2}+r^2d\theta^2\Big]\nonumber\\&-&a^2(d\chi^2+\sin^2\chi(d\lambda^2+\sin^2\lambda
d\psi^2)),\\G&=&\frac{2(1-a^2)}{R_1R_2^2g^2}(Vol_{R_1}(AdS_3)+Vol_{R_2}(S^3)),\\F^1&=&\frac{1-a^2}{g}\sin^2\chi\sin\lambda
d\lambda\wedge d\psi,\;F^2=\frac{1-a^2}{g}\sin\chi
d\chi\wedge d\lambda,\\F^3&=&\frac{1-a^2}{g}\sin\chi\sin\lambda
d\psi\wedge d\chi,\;e^{\sq\phi}=\frac{R_2^2g^2}{2(1-a^2)}.
\end{eqnarray}
Here $Vol_{R_i}(\mathcal{M})$ means the volume form of the manifold
$\mathcal{M}$ with radius of curvature $R_i$. Let us now introduce
coordinates $u$, $v$ according to 
\be
u=t+a\chi,\;\;\;v=t-a\chi,
\end{equation}
and take the Penrose limit along the null geodesic parametrised by
$u$. We obtain the metric
\be
R^{-2}_1d\bar{s}^2=dudv-\sin^2(u/2)g(\mathbb{R}^2)-a^2\sin^2(u/2a)g(\mathbb{R}^2).
\end{equation} 
Let us introduce coordinates $y^i$, $i=1,..,4$ such that the metric
takes the form
\be
R^{-2}_1d\bar{s}^2=dudv-\sum_1^4\frac{\sin^2(\lambda_iu)}{(2\lambda_i)^2}dy^idy^i,
\end{equation}
where
\be
\lambda_i=\Big\{\begin{array}{lcl}\frac{1}{2},\;\;\;i&=&1,2\\\frac{1}{2a},\;\;\;i&=&3,4.\end{array}
\end{equation}
Finally let us change coordinates to 
\be
x^-=R_1\Big(v+\frac{1}{4}\sum_iy^iy^i\frac{\sin(2\lambda_iu)}{2\lambda_i}\Big),\;\;\;x^+=\frac{R_1u}{2},\;\;\;x^i=R_1y^i\frac{\sin(\lambda_iu)}{2\lambda_i},
\end{equation}
so that the metric becomes
\be
d\bar{s}^2=2dx^+(dx^-+2\Big(\sum_i\Big(\frac{\lambda_i}{R_1}\Big)^2x^ix^i\Big)dx^+)-dx^idx^i.
\end{equation}
For the fluxes we obtain
\begin{eqnarray}
\bar{G}&=&\frac{2(1-a^2)}{R_1R_2^2g^2}dx^+\wedge (dx^1\wedge dx^2+dx^3\wedge
dx^4),\\\bar{F}^1&=&0,\\\bar{F}^2&=&\frac{1-a^2}{R_2^2g}\frac{dx^+}{\sqrt{(x^3)^2+(x^4)^2}}\wedge(x^3dx^3+x^4dx^4),\\\bar{F}^3&=&\frac{1-a^2}{R_2^2g}\frac{dx^+}{\sqrt{(x^3)^2+(x^4)^2}}\wedge(x^4dx^3-x^3dx^4).
\end{eqnarray}
In the limit $R_1\rightarrow\infty$, $R_2$ fixed, $G\rightarrow0$ and
the metric becomes that of $CW_4\times\mathbb{R}^2$. We could have
obtained this directly by taking the Penrose limit of the
$\mathbb{R}^{1,2}\times S^3$ solution. Finaly we note that the
vanishing of $\bar{F}^1$ is compatible with supersymmetry; after
taking the Penrose limit, the supersymmetry variation of the gauginos
is
\be
\d\bar{\lambda}^a=-\frac{1}{2\sq}e^{-\bar{\phi}/\sq}\bar{F}^a_{\m\v}\G^{\m\v}\e.
\end{equation}

\section{Conclusions}
We have found the most general supersymmetric ansatz for the six
dimensional chiral gauged $U(1)$ and $SU(2)$ theories, and explored the
geometrical structure of the solutions. Our results display both the
strengths and weaknesses of this general approach to finding
supersymmetric solutions. Because of the difficulties in solving
equations (\ref{u1base}) and (\ref{su2base}) in particular, we cannot
claim to have achieved the same degree of completeness in classifying
all supersymmetric solutions as was attained in \cite{tod} or
\cite{gaunt}, for example. However despite the fact that the theories
we have examined are 
considerably more complicated, they are still to some degree
tractable. Most encouragingly, we have demonstrated that the
nonabelian theory is at least no more intractable than the abelian
one, and one might hope that a similar approach applied to other
nonabelian gauged supergravities would be fruitful, and might allow
the construction of interesting new string/ M-theory solutions. We
have given a (rather implicit) classification of solutions of both
theories with enhanced supersymmetries. We
have also explored the Penrose limits of these gauged supergravities,
and shown that they yield pp-wave solutions of the ungauged theories.

Perhaps the greatest advantage in employing the G-structures approach
to the solution of supergravities is in the geometric insight one
obtains into the form of the supersymmetric solutions. For example in
the non-twisting case, finding solutions of our non-abelian gauged
supergravity in six dimensions essentially reduces to a problem in
pure four dimensional Riemanian geometry, namely the solution of
(\ref{su2base}) subject to the constraint (\ref{qquat}). Conceptually
this is an enormous simplification. However from a practical point of
view it is still a difficult problem, and we have only succeeded in
finding two explicit solutions.

Since the appearance of \cite{cve}, we know how to embed the
Salam-Sezgin model in string theory. It would be of 
interest to find the string theory realisation of the black string
solutions of the theories we have studied, and to study the
holography of the $AdS_3$ solutions. 

\section{Acknowledgements}
We would like to thank Gianguido Dall'Agata, Paul Davis, Juan Maldacena, Dario Martelli,
Carlos Nunez, James Sparks, Harvey Reall and in
particular Gary Gibbons for useful discussions. M. C. is supported by
EPSRC, Cambridge European Trust and Fondazione Angelo Della Riccia. OC is supported by a
National University of Ireland Travelling Studentship, EPSRC, and a
Freyer scholarship.

\end{document}